\documentclass[a4paper,10pt]{article}

\usepackage[english]{babel} 
\usepackage[latin2]{inputenc} 
\usepackage{graphicx} 
\usepackage{multirow} 
\usepackage[center]{caption}
\usepackage{geometry}
\usepackage{indentfirst}
\usepackage{txfonts}

\newenvironment{changemargin}[2]{%
  \begin{list}{}{%
    \setlength{\topsep}{0pt}%
      \setlength{\leftmargin}{#1}%
      \setlength{\rightmargin}{#2}%
      \setlength{\listparindent}{\parindent}%
      \setlength{\itemindent}{\parindent}%
      \setlength{\parsep}{\parskip}%
  }%
  \item[]}{\end{list}}

\usepackage{savesym}
\savesymbol{iint}
\savesymbol{iiint}
\savesymbol{iiiint}
\savesymbol{idotsint}
\savesymbol{openbox}

\usepackage{amsmath,amsthm}
\usepackage{amsfonts,amssymb}
\restoresymbol{TXF}{iint}
\restoresymbol{TXF}{iiint}
\restoresymbol{TXF}{iiiint}
\restoresymbol{TXF}{idotsint}
\restoresymbol{TXF}{openbox}

\usepackage{calc}
\usepackage{color}
\usepackage{dashrule}
\usepackage{dcolumn}
\usepackage{eurosym}

\usepackage{ifthen}
\newboolean{pdfflag}
\usepackage{ifpdf}
\usepackage{epstopdf}
\usepackage[T1]{fontenc}

\usepackage{multicol}
\usepackage{pifont}	
\usepackage{pstool}
\usepackage{relsize}
\usepackage{rotate}
\usepackage{setspace}
\usepackage[it]{subfigure}
\usepackage{verbatim}
\usepackage{epstopdf}

\newcommand{\Commentaire}[1]{}
\newcommand{\commentaire}[1]{}

\newcommand{\idr}[1]{_{\text{#1}}}
\newcommand{\edr}[1]{^{\text{#1}}}

\newcommand{\dd}{\text{d}}

\title{Comparison of the vibroacoustical characteristics of different pianos}
\author{Xavier Boutillon \\
   Laboratoire de Mécanique des Solides,\\
   École Polytechnique,\\
   91128 Palaiseau, France\\
   \texttt{boutillon@lms.polytechnique.fr}
   \and Kerem Ege \\
   Laboratoire Vibrations Acoustique,\\
   INSA-Lyon,\\
   25 bis avenue Jean Capelle,\\
   F-69621 Villeurbanne Cedex, France
   \and Stephen Paulello \\
   Stephen Paulello Piano Technologies,\\
   32 rue du sabotier, Hameau de Coquin,\\
   89140 Villethierry, France}
\date{}

\begin{document}
\twocolumn[
\maketitle
\begin{@twocolumnfalse}
\begin{changemargin}{1cm}{1cm}
On the basis of a recently proposed vibro-acoustical model of the piano soundboard (X. Boutillon and K. Ege, Vibroacoustics of the piano soundboard: reduced models, mobility synthesis, and acoustical radiation regime. \emph{submitted to the Journal of Sound and Vibration}, 2011.), we present several models for the coupling between the bridge and the ribbed plate of the soundboard. The models predict the modal density and the characteristic impedance at the bridge as a function of the frequency. Without parameter adjustment, the sub-structure model turns out to fit the experimental data with an excellent precision. The influence of the elastic parameters of wood is discussed. The model predictions are compared for pianos of different sizes and types.
\end{changemargin}\vspace{24pt}
\end{@twocolumnfalse}]

\section{Introduction}
\begin{figure}[ht!]
\begin{center}
\includegraphics[width = 0.48\textwidth]{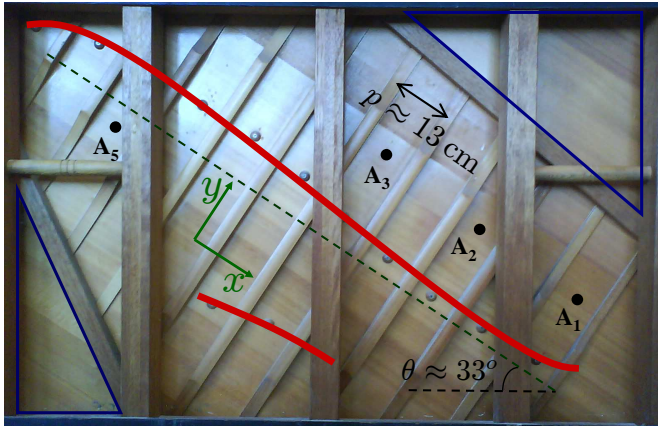}
\caption{Soundboard of the Atlas upright piano. Rib side with bridges superimposed as thick red lines. This upright soundboard include one ribbed zone and two cut-off corners (blue-delimited lower-left and upper-right triangles).}
\label{fig:pianoatlas}
\end{center}
\end{figure}

\begin{figure}[ht!]
\begin{center}
\includegraphics[width = 0.48\textwidth]{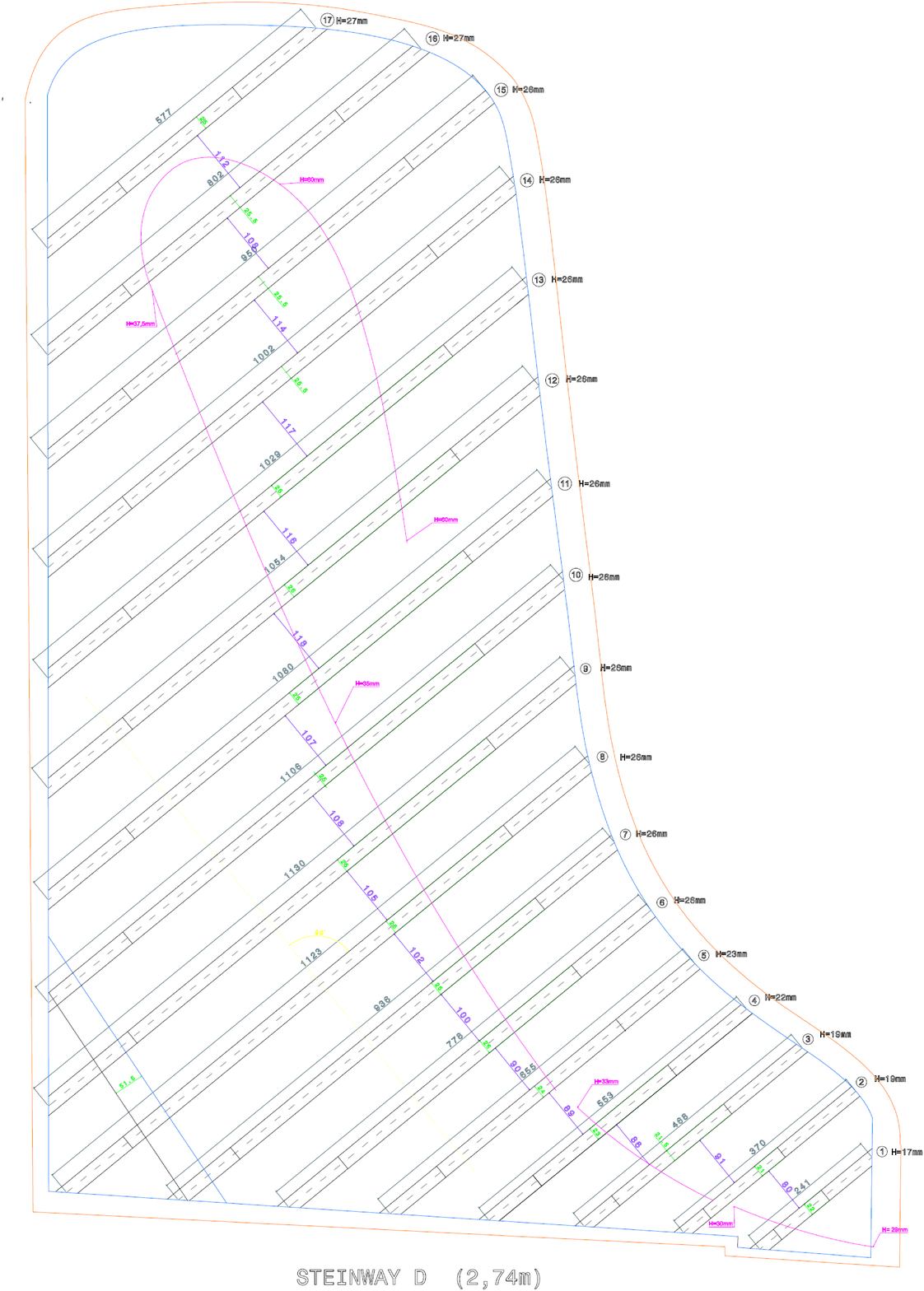}
\caption{Steinway model D. Geometry of the soundboard with 17 ribs, one cut-off corner, and a bridge.}
\label{fig:steinwayD}
\end{center}
\end{figure}

The piano soundboard (Figs.~\ref{fig:pianoatlas} and~\ref{fig:steinwayD}) is entirely made of wood. It consists in several parts: a panel on which is glued a slightly curved bar (the bridge), in the direction of the grain of the panel's wood. A series of thin, nearly parallel ribs are glued in the orthogonal direction. Eventually, thick bars isolate one or two cut-off corners which may exceptionally be ribbed themselves.

The grain of the main panel's wood defines the $x$-direction. The description of the soundboard relies on the following parameters:
\begin{itemize}
\item Material parameters: $\rho$, $E_x$ (or $c_x=\sqrt{E_x/\rho}$), $E_y$ (or $c_y=\sqrt{E_y/\rho}$), $\nu_{xy}$, $G_{xy}$ (or an orthotropy parameter $\gamma$, equal to one for elliptical orthotropy). 
\item Geometrical parameters: area $A$, geometry, boundary conditions (here, considered as clamped), dimensions of the various elements (wood panel, ribs, bridges, inter-rib spaces). The thickness $h$ of the wooden panel turns out to be an important element of the description.
\end{itemize}

It is assumed here that ribs are made of the same wood as the main panel: the Yong's modulus $E_r$ in their main direction is $E_x$.

\section{Vibratory regimes and models}

\begin{figure}[ht!]
\begin{center}
\includegraphics[width = 0.48\textwidth]{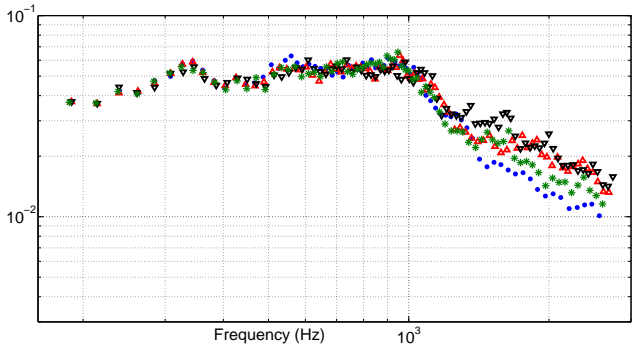}
\caption{Modal density of the Atlas piano soundboard. Dots: observed values at points \textbf{A}$_\mathbf{1}$ ({\color[rgb]{0,0,1}\tiny{$\bullet$}}), \textbf{A}$_\mathbf{2}$ ({\color[rgb]{1,0,0}$\blacktriangle$}), \textbf{A}$_\mathbf{3}$ ($\blacktriangledown$), and \textbf{A}$_\mathbf{5}$ ({\color[rgb]{0,0.5,0}\scriptsize{$\ast$}}), in Fig.\ref{fig:pianoatlas}. The estimated values are the reciprocal of the moving average of six successive modal spacings, reported at the mid-frequency of the whole interval.}
\label{fig:modaldensity}
\end{center}
\end{figure}

According to experimental modal analyses \cite{EGE2009_2,EGE2011} (see Fig.~\ref{fig:modaldensity}), the vibratory behaviour of a piano soundboard exhibits two distinct regimes:
\begin{itemize}
\item In low frequencies, the vibration extends over the whole soundboard, including the cut-off corners. The modal density is roughly constant and does not depend on the location of the point where the vibration is observed or generated.
\item Above a frequency \mbox{$f\idr{g}\approx 1.2$ kHz}, the modal density depends slightly on the point of observation and\\ strongly decreases with frequency. The vibration is located near the point $Q$ where it is observed or generated. More precisely, the vibration is confined between ribs which act as structural wave-guides.
\end{itemize}

Low-frequency models proposed below are based on orthotropic plate-elements representing large zones of the soundboard. In particular, ribs and the wood panel are considered altogether as a homogeneous plate. The high-frequency model is that of waves travelling in a structural wave-guide of width $p$, with $k_x=\dfrac{n\pi}{p}$. The vibration extends over three inter-rib regions: the one containing $Q$ and the two adjacent ones. The frequency limit $f\idr{g}$ between those two regimes (Tab.~\ref{tab:freqlim}) is obtained when $k_x$ reaches $\dfrac{\pi}{p}$ in the low-frequency model.

\begin{table}[ht!]
\begin{center}	
\begin{tabular}{|c|c|c|}
\hline
 \  & $p$~(cm)& $f\idr{g}$~(Hz) \\
\hline
Atlas & 13 & 1184  \\												 
\hline
Hohner & 11.2& 1589 \\
\hline
Schimmel & 12 & 1394 \\
\hline
Steinway model B & 12.2 & 1477 \\
\hline
Steinway model D & 12.7 & 1355 \\
\hline
\end{tabular}
\caption{Mean $p$ of the inter-rib widths for the different pianos and frequency limit $f\idr{g}$ between the low-frequency and the high-frequency regimes for average properties of wood.}
\label{tab:freqlim}
\end{center}
\end{table}

\section{Descriptive parameters}
For the string, the soundboard represents a mechanical impedance $Z(\omega)=\dfrac{F(\omega)}{V(\omega)}$. At a given location, the mobility $Y(\omega)=\dfrac{1}{Z(\omega)}$ can be computed as the sum of the mobilities of the normal modes of the structure. The description that is attempted here ignores the differences between locations of the string on the bridge. The models that are presented below do not predict damping which can be taken according to experiments or chosen more or less arbitrarily.

Given these hypotheses, the mechanical structures that compose the piano soundboard -- plates, bars, structural wave-guides -- are only characterised by the surfacic density $\mu=\rho h$ (in generic terms), one or several rigidities $D=\dfrac{E h^3}{12(1-\nu^2)}$ (\emph{idem}) or dynamical rigidities $d = \dfrac{D}{\mu}$, their length $L$ (for bars) or area $A$ (for plates), and their shape and boundary conditions. 

It can be shown that, except for the very first modes, this description is equivalent to:
\begin{itemize}
\item a modal density $n(f)$, depending mostly on $A$ (or $L$), $d$, and, for low frequencies only, on the shape and boundary conditions.
\item a characteristic impedance $Z\idr{c}$ (or mobility $Y\idr{c}$) which is the geometrical mean of $Z(f)$ (resp. $Y(f)$).
\end{itemize}

For a plate:
\begin{equation}
n\idr{p}(f)=\dfrac{A\idr{p}\,\zeta^{1/2}}{\pi\,d_x^{\,1/2}}\,F - n\idr{p,corr}(f)\qquad%
Y\idr{c,p}(f)=\dfrac{n\idr{p}}{r\,M\idr{p}}
\end{equation}
where $\zeta^2=E_x/E_y$, $F$ is a coefficient depending on the direction of orthotropy (typically $\pi/2$) and $n\idr{p,corr}(f)$ is a low-frequency correction depending on boundary conditions. The characteristic mobilities are given according to Skudrzyk's theory of the mean value~\cite{SKU1980}. In general, $r=4$, except when the plate is excited at one boundary, where it becomes $2$.

For a bar (such as the bridge):
\begin{equation}
n\idr{b}(f)\,=\,\dfrac{L\idr{b}}{d\idr{b}^{\,1/4}\,(2\pi\,f)^{1/2}}\qquad%
Y\idr{c,b}(f)=\dfrac{n\idr{p}}{r\,M\idr{p}}(1-j)
\end{equation}
In general, $r=4$, except when the bar is excited at one end, where it becomes $2$.

The case of a structural wave-guide needs a special discussion which cannot be included here; the result is the same as for a bar, with $r=2$ in general, $r=1$ when excited at one end.

\section{Homogenisation of a ribbed plate}
In the $y$-direction (weak direction of the panel's wood), the main zone of the soundboard (excluding cut-off corners) is stiffened by more or less regularly spaced ribs. The purpose of homogenisation is to derive the elastic properties of an orthotropic equivalent plate with similar mass, area, and boundary conditions as the main zone of the soundboard.

Following Berthaut \cite{BER2003,BER2004} and somewhat arbitrarily, we assume elliptical orthotropy for the equivalent plate. Thus, only two rigidities need to be considered, namely $D_x\edr{H}$ and $D_y\edr{H}$. Each rib (of width $a$) defines a cell extending between two mid-lines of adjacent inter-rib spaces. The rigidity of a portion of a cell of width $q$ and extending between $y$ and $y+\dd y$ is obtained by searching the position $H$ (in the $z$-direction, orthogonal to the soundboard plane) of the neutral line that minimises the composite rigidity of the plate associated with the rib of height $\beta$. It comes:
\begin{align}
H(y) &= \dfrac{-q E_y h^2 + a E\idr{r}\beta^2}{2\left(q E_y h + a E\idr{r}\beta\right)}\\
D_y(y) &= \dfrac{E_y}{3}(h^3 + 3 h^2 H + 3 h H^2)+\dfrac{E\idr{r}a}{3q}(\beta^3 - 3 H\beta^2 + 3 H^2\beta)
\end{align}

Since ribs are slightly irregularly spaced along the $x$-direc-tion (cell have different widths $q(x)$) and since each rib has a varying height $\beta(y)$ along the $y$-direction, we adopt the approximation  that $1/D_{x,y}\edr{H}$ are the average flexibilities (inverse of rigidities) in each direction. The computation has been made numerically, on the basis of the geometry of each rib and inter-rib space.

\section{Models for the association of the\\ bridge and the ribbed plate}
How to describe the association between the ribbed plate and the bridge has been debated for long \cite{LIE1979,CON1996_2,EGE2011_2}. Three solutions are presented here. They are compared when applied on the piano labelled "Atlas", of which we know simultaneously the detailed geometry and the results of an experimental modal analysis.
\begin{figure}[ht!]
\begin{center}
\includegraphics[width = 0.48\textwidth]{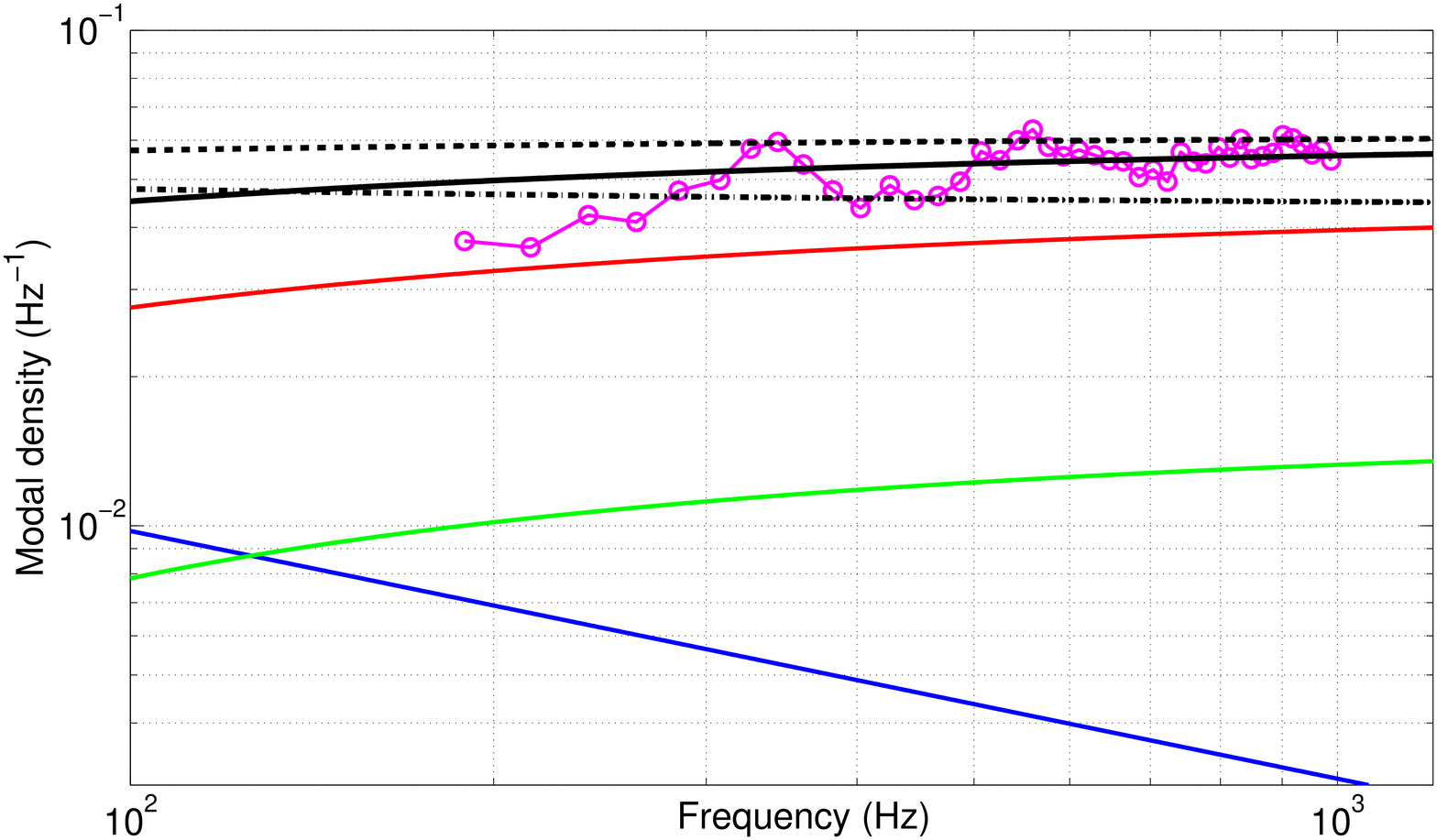}
\caption[aaa]{Modal densities in the Atlas upright piano, for different models.\\
Solid blue line: bridge (mapple). Solid green line: cutoff corners (Norway spruce). Solid red line: sum of the modal densities of the sub-plates (Norway spruce) separated by the bridge (homogenised equivalent plate). Solid black line: total of the previous modal densities (sub-structure model).\\
Dashed line: modal density of the whole soundboard according to the first approximation proposed by Skudrzyk (independant plate and bridge).\\
Dash-dotted line: modal density of the whole soundboard according to the second approximation proposed by Skudrzyk (see text).}
\label{fig:densite_modale_models}
\end{center}
\end{figure}
\begin{figure}[ht!]
\begin{center}
\includegraphics[width = 0.48\textwidth]{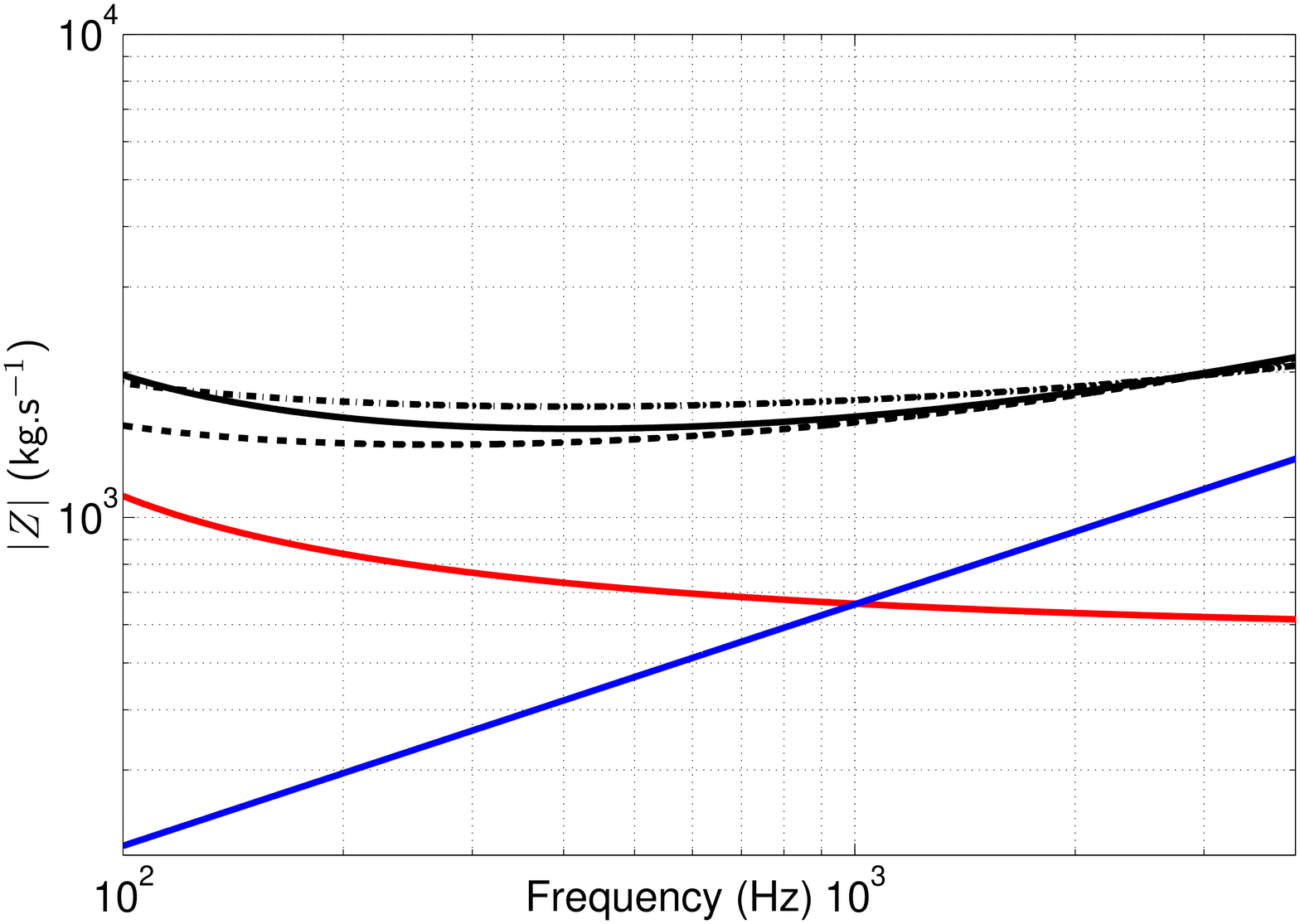}
\caption[aaa]{Characteristic impedances of the Atlas upright piano, for different models.\\
Solid blue line: bridge (mapple). Solid red line: sum of the characteristic impedances of the sub-plates (Norway spruce) separated by the bridge (homogenised equivalent plate). Solid black line: total of the previous characteristic impedances (sub-structure model).\\
Dashed line: characteristic impedance of the whole soundboard according to the first approximation proposed by Skudrzyk (independant plate and bridge).\\
Dash-dotted line: characteristic impedance of the whole soundboard according to the second approximation proposed by Skudrzyk (see text).}
\label{fig:impedances_models}
\end{center}
\end{figure}

According to Skudrzyk \cite{SKU1980}, a simple approximation consists in considering the plate and the bridge as uncoupled. It follows that the modal densities simply add and that the characteristic impedances of each element add as well. Skudrzyk's comment is that the resulting error is small because this approximation generates two errors that partly compensate each other. The results are given by the dashed black lines in Figs.~\ref{fig:densite_modale_models} (modal density) and \ref{fig:impedances_models} (characteristic\\ impedance).

Skudrzyk presents a supposedly better approximation for describing the association of a single bar with a plate: the dynamics in the bridge direction (here: $Ox$) is ruled by the bridge, which must be considered as mass-loaded by the plate. On the other hand, the plate must be considered as stiffened by the bridge. The modal densities and characteristic impedances of these modified elements must then be added. Based on our understanding of Skudrzyk's expressions, results are given by the dash-dotted black lines in Figs.~\ref{fig:densite_modale_models} (modal density) and \ref{fig:impedances_models} (characteristic impedance). As a matter of fact, the match with experimental data is not better than the previous, more simple approximation.

A third interpretation of the coupling between the bridge and the ribbed plate can be given in terms of sub-structures. Since the bridge extends over almost the whole soundboard, we have considered that it splits the main zone of the soundboard in two plates, and represents an quasi-boundary condition for each of the two (sub-)plates. Due to the contrast in stiffness between the bridge and the equivalent plate, we assumed a clamped boundary condition. The results are given in solid black lines Figs.~\ref{fig:densite_modale_models} (modal density) and \ref{fig:impedances_models} (characteristic impedance). Since, in our view, this model is better grounded and yields results which better fit experimental findings, it is adopted in the rest of the article. It should be noticed that the different models yield much closer values for the characteristic impedance than for the modal densities.

\section{Influence of the characteristics\\ of wood}
\begin{figure}[ht!]
\begin{center}
\includegraphics[width = 0.48\textwidth]{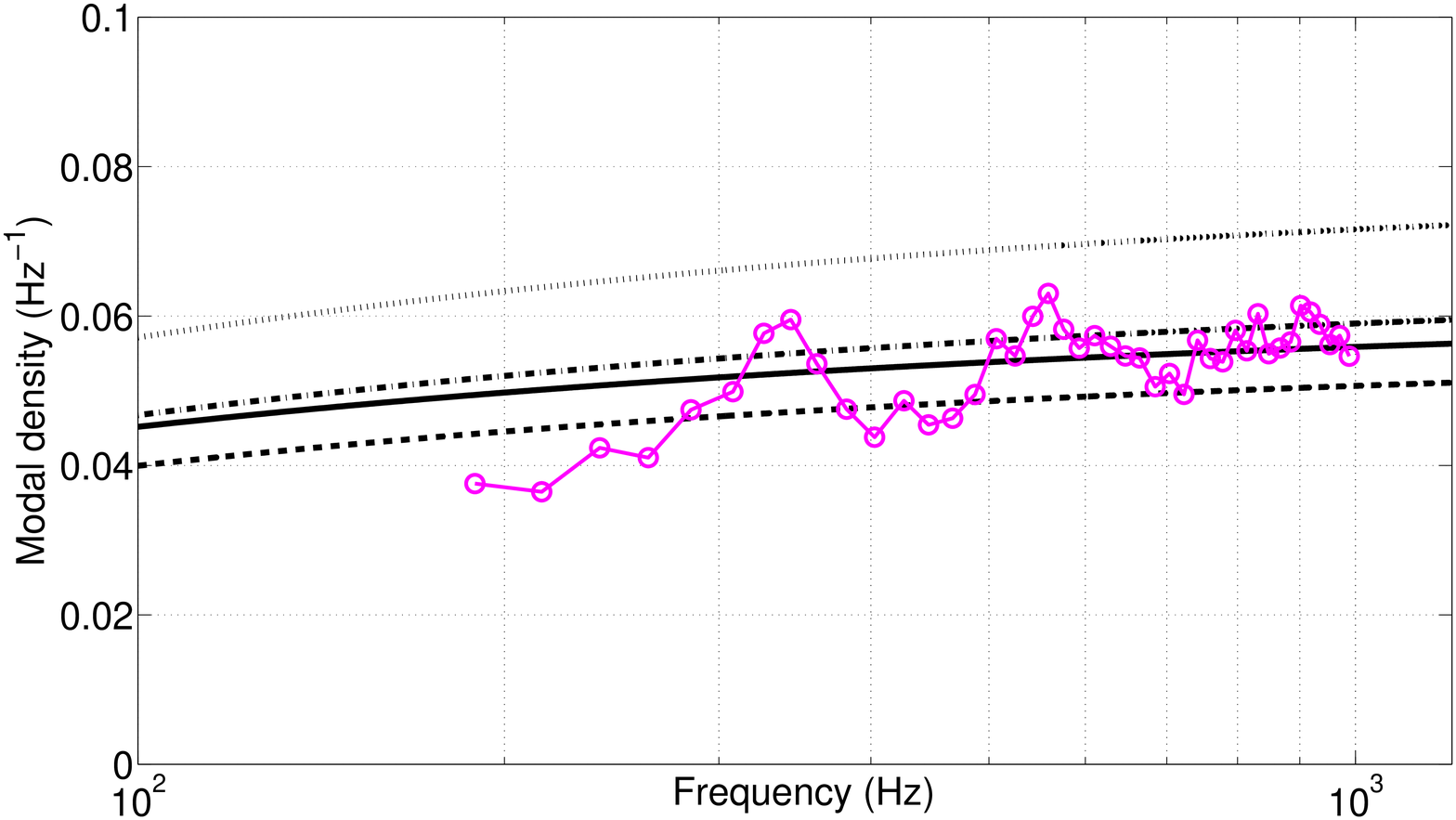}
\caption[aa]{Modal densities of the Atlas upright piano obtained for different elastic parameters of spruce.\\
Solid line: Norway spruce (\mbox{$\rho=440$ kg/m$^3$}, \mbox{$E_x=$ 15.8 GPa} (corresponding to \mbox{$c_x=$ 6000 m/s}), \mbox{$E_y =$ 0.85 GPa} (corresponding to \mbox{$c_y =$ 1400 m/s})).\\
Dash-dotted line: average spruce (\mbox{$\rho=380$ kg/m$^3$}, \mbox{$E_x=$ 11.5 GPa} (corresponding to \mbox{$c_x=$ 5500 m/s}), \mbox{$E_y =$ 0.74 GPa} (corresponding to \mbox{$c_y =$ 1400 m/s})).\\
Dotted line: mediocre spruce (\mbox{$\rho=400$ kg/m$^3$}, \mbox{$E_x=$ 8.8 GPa} (corresponding to \mbox{$c_x=$ 5000 m/s}), \mbox{$E_y =$ 0.35 GPa} (corresponding to \mbox{$c_y =$ 1000 m/s})).\\
Dashed line: excellent spruce (\mbox{$\rho=350$ kg/m$^3$}, \mbox{$E_x=$ 12.6 GPa} (corresponding to \mbox{$c_x=$ 6000 m/s}), \mbox{$E_y =$ 1.13 GPa} (corresponding to \mbox{$c_y =$ 1800 m/s})).\\
Circles: experimental determinations.}
\label{fig:densite_modale_elasticparameters}
\end{center}
\end{figure}

\begin{figure}[ht!]
\begin{center}
\includegraphics[width = 0.48\textwidth]{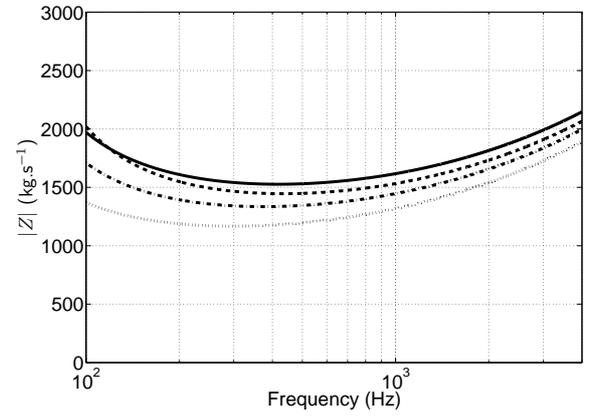}
\caption{Characteristic impedances of the Atlas upright piano obtained for different elastic parameters of spruce. See Fig.~\ref{fig:densite_modale_elasticparameters}.}
\label{fig:impedance_elasticparameters}
\end{center}
\end{figure}

It is very difficult to know precisely what are the elastic properties of woods in a given piano. In the results presented above, we have retained values given by the literature \cite{HEA1948,HAI1979} for Norway spruce: \mbox{$\rho=440$ kg/m$^3$}, \mbox{$E_x=$ 15.8 GPa} (corresponding to \mbox{$c_x=$ 6000 m/s}), \mbox{$E_y =$ 0.85 GPa} (corresponding to \mbox{$c_y =$ 1400 m/s}). The influence of the wood quality are presented in Figs.~\ref{fig:densite_modale_elasticparameters} (modal density) and~\ref{fig:impedance_elasticparameters} (characteristic impedance) for four sets of values corresponding to Norway spruce, to average values (according to the literature and to collected experience of one of us), to a mediocre wood, and to an excellent wood. It is clear that the precision of the fit is subject to a correct knowledge of the wood. 


\begin{figure}[ht!]
\begin{center}
\includegraphics[width = 0.48\textwidth]{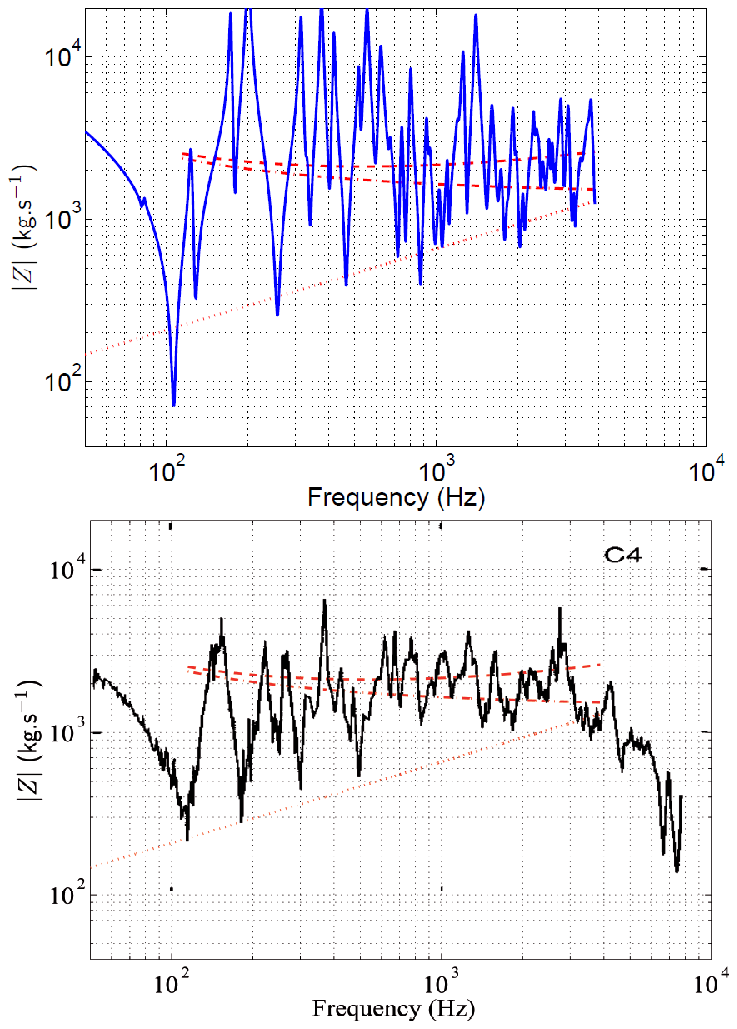}
\caption[]{Red lines (identical in both frames): characteristic impedance reconstructed for the piano measured by Giordano~\cite{GIO1998} according to the second Skudrzyk's model (same as in Fig.~\ref{fig:impedances_models}).\\
Upper frame: synthesised impedance. Lower frame: Measured impedance, after Giordano~\cite{GIO1998}.}
\label{fig:figures_impedances_giordano}
\end{center}
\end{figure}

\section{Application to different pianos}
\begin{figure}[ht!]
\begin{center}
\includegraphics[width = 0.48\textwidth]{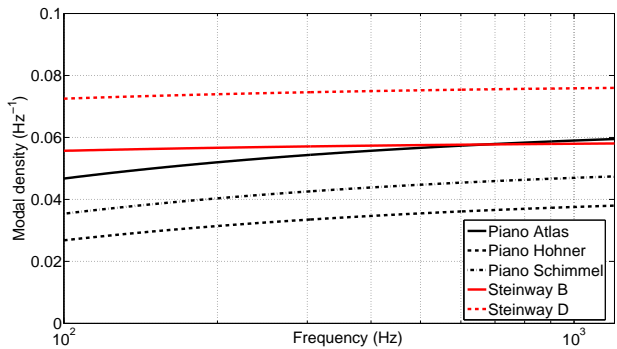}
\caption{Modal densities of three upright pianos, and two grand pianos (values obtained for average characteristics of spruce).}
\label{fig:compar_n}
\end{center}
\end{figure}

\begin{figure}[ht!]
\begin{center}
\includegraphics[width = 0.48\textwidth]{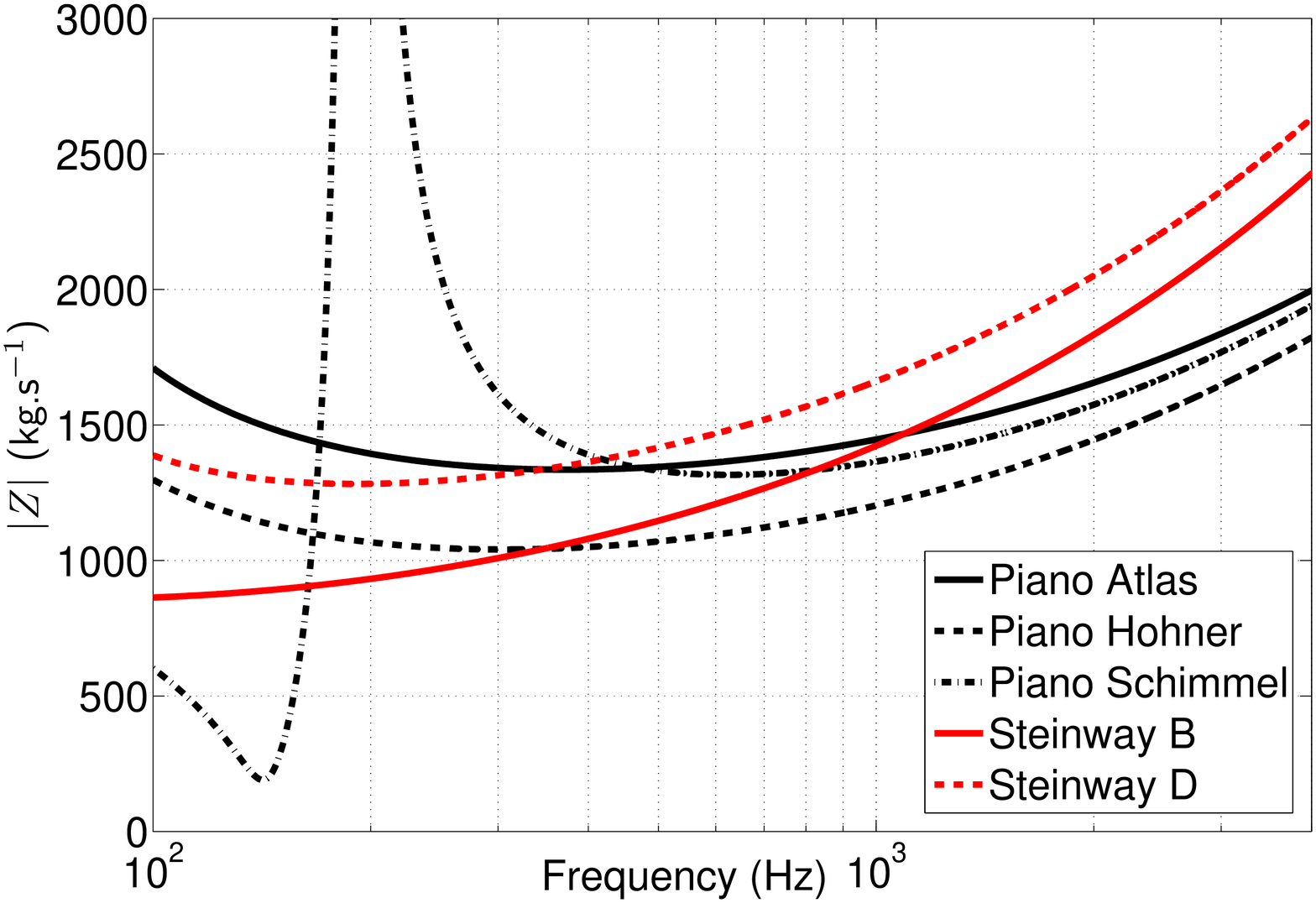}
\caption{Characteristic impedances of three upright pianos, and two grand pianos (values obtained for average characteristics of spruce). The low-frequency strong variation for the Schimmel upright is an artefact of the model.}
\label{fig:compar_imp}
\end{center}
\end{figure}

The sub-structure model has been applied to different pianos: three uprights (Atlas, Hohner, Schimmel, respectively of height 120, 110, and \mbox{120 cm}) and two grands (Steinway B and Steinway D). One Bösendorfer (Imperial prototype) is currently under investigation but not reported here. The predicted modal densities, up to $f\idr{g}$ are presented in Fig.~\ref{fig:compar_n} and the predicted characteristic impedances in Fig.~\ref{fig:compar_imp}. The most striking feature of these figures is that these parameters do not differ considerably between pianos. This is remarkable considering that, for example, a \mbox{1 mm} variation in the thickness of the soundboard\footnote{In the Steinway B and D, the thickness of the wood panel varies between 9 mm in the centre to 6 mm at the rim.} causes a variation in modal density and impedance that is of the same order of magnitude as the differences between pianos. The same can be observed with regard to the variation in the characteristics of wood. One may therefore conclude that the modal density and characteristic impedance are typical of the identity of the piano instrument, as such, at least in the present days.

An interesting difference between uprights and grands is that grands have a larger modal density \emph{and} a larger characteristic impedance. The larger modal density can be seen as a natural consequence of the increase in size. However, for a homogeneous plate, the standard relationship between $n$ and $Y\idr{c}$ would lead to a variation of $n$ and $Z\idr{c}$ in opposite directions. One may conclude that a similar variation is only attained by a careful geometrical design.

Observing the details of the geometries reveals rib spacing is slightly irregular \emph{for all pianos}. We have interpreted this elsewhere \cite{EGE2011_2} as a way to \emph{localise} the vibration in high frequencies. Space is missing here to report on this in more details.

It can also be observed that the height of the ribs, the proportion between the high part of the rib and each of their end parts, sometimes their width, systematically vary from bass to treble. On one of the uprights, even the rib height of the central part varies, and this differently along each rib. This indicates a careful adjustment of the \emph{local} mechanical (and therefore, vibratory) properties. All these are ignored by the present global, average model. This variation of the impedance as a function of the pitch needs to be examined in order to account for the quality of each piano model, in other words, for the rendered match between the acoustical properties of a note and its pitch.

\section*{Acknowledgments}
We heartly thank Claire Pichet for her careful technical drawings and the dimension report of the grand pianos; this made the input of the geometry into the programs considerably easier.
\bibliographystyle{unsrt}
\bibliography{PianoComparison}

\end{document}